\documentstyle[epsf]{article}
\def\1ad{\mbox{\normalsize $^1$}}
\def\2ad{\mbox{\normalsize $^2$}}
\def\3ad{\mbox{\normalsize $^3$}}
\def\4ad{\mbox{\normalsize $^4$}}
\def\5ad{\mbox{\normalsize $^5$}}
\def\6ad{\mbox{\normalsize $^6$}}
\def\7ad{\mbox{\normalsize $^7$}}
\def\8ad{\mbox{\normalsize $^8$}}
\def\makefront{\vspace*{1cm}
\hfill CERN-TH/99-85\vskip -2pt
\hfill hep-th/9903232\vskip -5pt
\begin{center}
\def\newtitleline{\\ \vskip 0pt}
{\Large\bf\titleline}\\
\vskip 1truecm
{\large\authors}\\
\vskip 5truemm
\addresses
\end{center}
\vskip .5truecm
{\bf Abstract:}
\abstracttext
\vskip 1truecm}
\catcode`\@=11
\def\mref#1{\ifx\und@fined#1
{need to supply reference \string#1.}
\else #1 \fi}
\catcode`\@=12 %
\let\lref=\def

\newcommand\cA{{\cal A}} 
 \newcommand\cD{{\cal D}}
\newcommand\cE{{\cal E}}
 
\newcommand\cH{{\cal H}} 
 
\newcommand\cL{{\cal L}}

\newcommand\inbar{\vrule height1.5ex width.4pt depth0pt}
\newcommand\IC{\relax\,\hbox{$\inbar\kern-.3em{\rm C}$}}
\newcommand\IQ{\relax\,\hbox{$\inbar\kern-.3em{\rm Q}$}}
\newcommand\IR{\relax{\rm I\kern-.18em R}}
\newcommand\IP{\relax{\rm I\kern-.18em P}}
\def\ZZ{\relax{\sf Z\kern-.4em Z}}
\newcommand{\beq}{\begin{equation}}
\newcommand{\eeq}{\end{equation}}
\newcommand{\eel}[1]{\label{#1}\end{equation}}
\newcommand{\bea}{\begin{eqnarray}}
\newcommand{\eea}{\end{eqnarray}}
\newcommand{\eeal}[1]{\label{#1}\end{eqnarray}}
\newcommand{\beac}{\begin{equation}\begin{array}{rcl}}
\newcommand{\eeacn}[1]{\end{array}\label{#1}\end{equation}}

\newcommand{\eq}[1]{(\ref{#1})}

\newcommand\Coeff[2]{{#1\over #2}}
\newcommand\coeff[2]{\relax{\textstyle {#1 \over #2}}\displaystyle}

\newcommand\attac[1]{\Bigl\vert
{\phantom{X}\atop{{\rm\scriptstyle #1}}\phantom{X}}}

\newcommand\figinsert[4]
{\begin{figure}[htb]
\newcommand\figsize{#3}
\epsfysize\figsize
\centerline{\epsffile{#4}}
\vskip -.2cm
\caption{\leftskip 1pc \rightskip 1pc
\baselineskip=10pt plus 2pt minus 0pt {\sl {#2}}}
\label{fig:#1}
\end{figure}}
\newcommand\taueff{\tau_{{\rm eff}}}

\newcommand\tr{{\rm tr}}
\newcommand\LCS{\cL^{\rm (CS)}}
\newcommand\Aroof{{\widehat{\cA}}}
\newcommand\Lroof{{\widehat{\cL}}}
\newcommand\taus{\tau_S}
\def\Cp#1{C^{(#1)}}

\begin {document}
\def\titleline{
On the Anomalous and Global Interactions
\newtitleline
of Kodaira 7-Planes\footnote
{Contribution to the proceedings of the 32nd International
Symposium on the Theory of Elementary Particles, Buckow 1998,
to appear in "Fortschritte der Physik".}
}
\def\authors{
W.\ Lerche  and S.\ Stieberger
}
\def\addresses{
Theoretical Physics Division, CERN\\
1211 Geneva 23, Switzerland
}
\def\abstracttext{

We review interactions between certain 7-planes which are composed out
of mutually non-local $(p,q)$ 7-branes and which correspond to
specific Kodaira singularities. We discuss in particular how to
compute certain moduli-dependent terms in the effective action. These
do not only probe the local Chern-Simons terms on the
world-volumina, but also certain global aspects of plane interactions
that can be attributed to ``torsion'' (or $\ZZ_N$-valued) $D$-brane
charges.

}
\large
\makefront

\section{Introduction}

When studying interactions between $D$-branes, one usually considers
idealized situations where one focuses one a single pair of (possibly
stacks of) branes. However, for making quantitative  tests of string
dualities, the global structure of $D$-branes interactions become
important as well -- that is, the influence of all
the other branes that are usually considered as far away. The purpose
of this contribution is to elucidate some aspects of such ``global
interactions'', reviewing part of ref.\ \cite{WLSS} but also slightly
expanding upon some of the points.

The theory we are concretely interested in here is $F$-theory
compactification on $K3$, which amounts to type IIB compactification on
$\IP^1$ with 24 7-branes \cite{Fth}. This theory is dual to the
heterotic string  on $T^2$, where certain anomaly-related pieces of
the effective action can be computed exactly at one-loop order. This
fact provides an ideal framework for studying non-trivial brane
interactions; in the past, it has been very successfully applied to
study brane interactions in type I and matrix strings (see eg.,
\cite{BFKOV,ko,typeI}).

Obviously, one cannot expect to compute any given piece of the
effective lagrangian exactly; it is just very special couplings, namely
typically those which are anomaly-related and/or to which only
BPS-states contribute, that are amenable to an exact treatment. In the
present situation with 16 supercharges in eight dimensions, the
canonical BPS-saturated amplitudes \cite{WL,BK} involve four external
gauge bosons (and/or gravitons). Supersymmetry relates parity even
($i\xi\,F^n$) and parity odd (${\theta\over2\pi} F \wedge F\wedge ..
F$) sectors, and therefore one can conveniently combine the theta-angle
and the coupling constant $\xi$ into one complex coupling, $\taueff$.
In particular, when compactifying the heterotic string on $T^2$, the
effective couplings $ \taueff(T,U) \equiv i\xi(T,U)+ \Coeff1{2\pi}
\theta(T,U)$ become highly non-trivial functions depending on the usual
torus moduli $T$ and $U$.  As mentioned before, in the heterotic string
picture the couplings are exact at one-loop order, and are in fact
directly related \cite{WL} to the heterotic elliptic genus
$\Aroof_{}(F,R,q)$ \cite{ellg}. Schematically:
\beq
{\rm Re}[\taueff(T,U)]\, F\wedge ...F\ \,
\sim\ \int {d^2\tau\over {\tau_2}}
Z_{(2,2)}(T,U,q,\bar q)\Aroof_{}(F,R,q) \attac{8-form}\ ,
\eel{ellgen}
where $Z_{(2,2)}$ is the partition function of the two-torus $T^2$ (we
will switch off the Wilson lines, or set them to certain constant
values). The evaluation of such integrals has been discussed at length
in \cite{HM,ko,WLSS,LSWA}, and results in certain Borcherds-type of
modular functions. The resulting expressions have been interpreted in
terms of type I string language \cite{BFKOV,ko,typeI}, in which one has
been able to identify both perturbative and non-perturbative
($D$-instanton) contributions. On the other hand, an interpretation in
terms of $F$-theory geometry was given in  refs.\
\cite{WLSS,LSWA,LSWB}, and this is what we will --partially-- review
below.

\section{Chern-Simons Couplings on 7-Planes}

The issue is to compute the functions $\taueff(T,U)$ \eq{ellgen} via
7-brane interactions. While in general very complicated, these
interactions in the present context reasonably tractable because of
their special anomaly related, parity-odd structure. They arise from
the Chern-Simons terms on the world-volumina of the 7-branes, via the
exchange of $RR$ antisymmetric tensor fields $\Cp p$. For a single
$D$-brane, the relevant tree level couplings look \cite{GHM} (for
trivial normal bundle):
\beq
\LCS_{D7}\ =\ C\wedge e^{-2iF}\wedge \sqrt{\Aroof(R)} \attac{8-form}\ ,
\eel{Lcs}
where $C\equiv\oplus_{k=0}^4 \Cp {2k}$ is the formal sum over all $RR$
forms, and $\Aroof(R)$ is the Dirac genus.

Note that due to the generic mutual non-locality of the 24 $(p,q)$
7-branes, it is a priori not clear how to add up these terms, and much
less, how to determine what effective interactions they induce.
However, what we can do is simply to restrict to sub-moduli spaces
where the 24 branes combine into certain objects (``7-planes'') that
are all mutually local, ie., have commuting monodromies.
Particularly simple are the sub-cases where the monodromies not only
commute, but are also of finite order. These correspond to theories
where the 7-brane charge is cancelled locally, such that the type IIB
string coupling $\taus$, is constant over the $\IP^1$ base (the
``$z$-plane''). There are three such possibilities for splitting up the
24 7-branes \cite{Sen,KDSM}:

\noindent {\bf ii)} $\taus=i$: into eight $\cH_1$-planes
(five remaining moduli)

\noindent {\bf ii)} $\taus=\rho\equiv e^{2 \pi i/3}$: into twelve
$\cH_0$-planes (nine remaining moduli)

\noindent {\bf iii)} $\taus=$arbitrary constant: into four
$\cD_4$-planes
(one remaining modulus; this branch intersects branches ii) and iii)).

Here $\cH_n$ denotes 7-planes associated with the Kodaira elliptic
singularity types \cite{Kod} of the same name; they carry gauge
symmetries of type $A_n$ ($A_0\sim U(1)$), respectively. By further
specialization, one can have some or all of the $\cH_n$ branes combine
into planes with larger gauge symmetries; the various possibilities are
listed in the following table.

{\vbox{{
$$
\vbox{\offinterlineskip\tabskip=0pt
\halign{\strut
\vrule
{}~$#$
\vrule
& ~$#$~
& ~$#$~
& ~$#$~
& ~$#$~
& ~$#$~
& ~$#$~
& ~$#$~
&\vrule#
\cr
\noalign{\hrule}
{\rm Kodaira\ type}\to
&
\cH_0
&
\cH_1
&
\cH_2
&
\cD_4
&
\cE_6
&
\cE_7
&
\cE_8
&
\cr
\noalign{\hrule}
{\rm Composition}
&
\cH_0
&
\cH_1
&
{\cH_0}^2
&
{\cH_0}^3,{\cH_1}^2
&
{\cH_0}^4
&
{\cH_1}^3
&
{\cH_0}^5
&
\cr
{\rm Monodromy}
&
(ST)^{-1}
&
(S)^{-1}
&
(ST)^{-2}
&
(S)^{2}\equiv -{\bf 1}
&
(ST)^{2}
&
S
&
ST
&
\cr
{\rm Discrete\ Charge}
&
\ZZ_6
&
\ZZ_4
&
\ZZ_3
&
\ZZ_2
&
\ZZ_3
&
\ZZ_4
&
\ZZ_6
&
\cr
\taus
&
\rho
&
i
&
\rho
&
{\rm any}
&
\rho
&
i
&
\rho
&
\cr
\noalign{\hrule}}
\hrule}
$$
\vskip-10pt
\noindent{\bf Table 1:}
{\sl
List of 7-planes with finite order monodromies in the $z$-plane, which
do not carry net ($\ZZ$-valued) $D$-brane charge. They support gauge
fields on their world-volumina corresponding to the respective Kodaira
singularity type ($\cH_n\sim A_n$). We also list their composition in
terms of basic building blocks, as well as the associated constant type
IIB string coupling, $\taus$. ($S,T$ denote the standard generators of
the $S$-duality group, $SL(2,\ZZ)$.)
}
\vskip10pt}}}

Because of the mutual locality, the anomalous couplings can be very
simply determined. For this we recall  that a $\cD_4$-plane can be
viewed as being composed out of four $D7$-branes plus one orientifold
plane, which all are mutually local. Since $\LCS_{O7}\ =\ -4C\wedge
\sqrt{\Lroof(R)}|_{\rm{8-form}}$ \cite{DJM,MSM,OHirz}, where
$\Lroof(R)$ is
the Hirzebruch genus, we thus have:
\bea
\LCS_{\cD_4}\ &=&\
C\wedge\left[\tr\big(e^{-2iF}\big)\wedge\sqrt{\Aroof(R)}-
4\sqrt{\Lroof(R)} \right]\attac{8-form}\cr
&=& \Cp4\!\wedge\Big(\coeff12 R^2-2\tr F^2\Big)+
\Cp0\!\wedge \Big(\coeff23\tr F^4-\coeff1{12}\tr F^2 \tr R^2
+\coeff1{192}(\tr R^2)^2+\coeff1{48}\tr R^4 \Big)
\cr
&=:& 3 \LCS_{\cH_0} \ =:\ 2\LCS_{\cH_1}\ ,
\eeal{D4CS}
where in the last line the gauge field traces follow implicitly from
the decomposition $SO(8)\to U(1)$ or $SU(2)$, respectively. Indeed,
summing over all world-volumina exactly reproduces the (eight
dimensional remainder of the) Green-Schwarz term of the heterotic
string, ${\cal L}^{(GS)}=\Cp6\!\!\wedge2(R^2\!-\!\tr
{F_{SO(32)}}^2)\!+\!\Cp2\!\wedge\! X_8({F_{SO(32)},R})$.

\section{Geometric Interactions}

What we are interested in are the non-trivial interactions between the
planes, which should ultimately reproduce the coupling functions
$\taueff(T,U)$ that were schematically computed in \eq{ellgen}.  The
primary perturbative contributions will arise from strings stretched
between the various planes, which amounts to $\Cp p$ tensor field
exchange between individual planes. The effective interaction will thus
depend on the distances between the various 7-planes in the $z$-plane.
More specifically, the closed string exchange that contributes to the
maximal number of wedge products of field strengths is in the odd $RR$
sector, and is proportional to the Green's function $\Delta$ of a
scalar field on the $z$-plane. Schematically,\footnote {For a more
precise discussion, see eg.\ \cite{MSM}.}
\bea \Big\langle\Cp
p_{m_1\dots m_p}(z_1)\Cp{8-p}_{n_1\dots n_{8-p}}(z_2)
\Big\rangle_{{RR^-}}&=&  \epsilon^{m_1\dots m_pn_1\dots
n_{8-p}}\Delta(z_1,z_2)\ ,\cr {\rm where}\ \
\Delta(z_1,z_2)&=&\ln(z_1-z_2)+{\rm finite}\ .
\eeal{greensf}
The dependence of the plane locations $z_i$ on $T,U$ is
complicated, but can be obtained via the $K3$ mirror map, as explained
in
ref.~\cite{WLSS,LSWA}.

However, in order to obtain functionally exact results, we need to know
the full Green's functions that probe the global structure of the
$z$-plane, and not just their leading singular behavior. This is in
general a complicated problem, but in our setup, where we consider only
planes with finite order monodromies, there is a natural geometric
answer \cite{WLSS,LSWA}.  This is because the branch cuts in the
$z$-plane are of finite order, and therefore the geometry is
effectively the one of Riemann surfaces $\Sigma$ -- see Fig.1 (indeed
the geodesic lengths of strings stretching in the $z$-plane, given by
$\int_{z_1}^{z_2}dz\prod_i (z-z_i)^{-1/12}$, turn out to be given by
the periods of $\Sigma$).

\goodbreak
\figinsert{cover}{
Lift of the $z$--plane to a covering Riemann surface. Shown is
here the situation with two $\cE_6$ and two $\cH_2$ planes, which
correspond to $\ZZ_3$ twist fields and anti-twist fields, respectively,
located at the branch points of a genus two curve $\Sigma_2$.
We also show a closed string trajectory that contributes to the
coupling $\tr {F_{SU(3)}}^2\tr {F_{SU(3)'}}^2$, and which corresponds
to a $1/3$--period on $\Sigma_2$.}{1.9in}{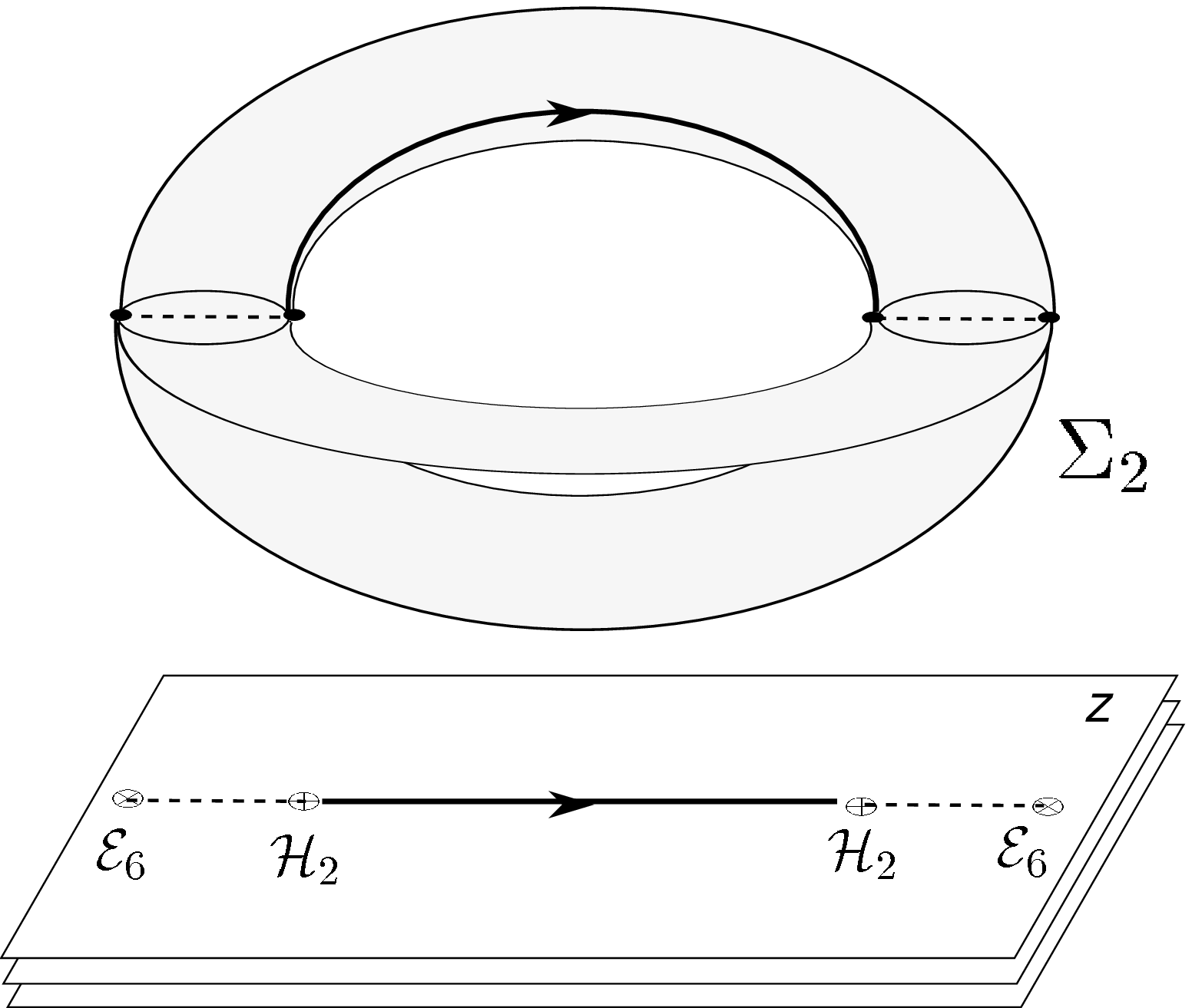}

Accordingly, canonical Green's functions are given by the ones of
scalar fields on the covering surfaces $\Sigma$: those are known to be
given by the logarithm of the ``prime form''. However, it turns out
\cite{LSWA} that these canonical Green's functions on $\Sigma$ to not
capture the full story; they capture only the perturbative piece of the
interactions. In fact, a Green's function is ambiguous up to the
addition of finite pieces, and indeed there are in general further
non-singular non-perturbative corrections to the naive Green's
functions, arising from wrapping string world-sheets on $\Sigma$. It
was shown in \cite{LSWA} how the non-perturbative Green's functions
between pairs of branch points can be computed exactly by solving
certain differential equations.

Due to lack of space, we will consider in the following mainly the
interactions of model iii), which is Sen's model with $SO(8)^4$ gauge
symmetry. Here four $\cD_4$ 7-branes create two $\ZZ_2$ branch cuts, so
that the covering space is just a torus, $\Sigma=T^2$. The situation is
especially simple also because the non-perturbative, $U$-dependent
contributions factor out, so that the naive Green's functions on the
torus give the full result (as far as the $T$-dependence is
concerned\footnote{This is reflected by the fact that the $z$-plane
geometry does not depend on the modulus $U$, which plays the role of
$\taus$. Non-trivial $U$-dependence is induced by $D$-instantons, to
the effect that $\Cp0$ in \eq{D4CS} is renormalized into
const$\times\ln|\eta(U)|^4$. This can be seen by comparing to the
$U$-dependence as computed on the heterotic side \cite{WLSS}.}). More
explicitly, the (harmonic parts of the) Green's functions between the
four branch points are given by
\beq
\Delta_{12}=\Delta_{34} =
\ln\Big[{\theta_2(2T)\over\theta_1'(2T)}\Big],\ \
\Delta_{13}=\Delta_{24} =
\ln\Big[{\theta_3(2T)\over\theta_1'(2T)}\Big],\ \
\Delta_{14}=\Delta_{23} =
\ln\Big[{\theta_4(2T)\over\theta_1'(2T)}\Big].
\eel{torusgreen}
Integrating out $\Cp4$ exchange between the Chern-Simons couplings
in \eq{D4CS} therefore immediately yields the following terms
in the effective action:
\beq
\cL^{{\rm eff}}\ \sim\ \sum_{i<j}^4{\rm Re}[\Delta_{ij}(T)]
\tr F_{SO(8)_i}^2\wedge \tr F_{SO(8)_j}^2\ .
\eel{FiFj}
These can be seen \cite{WLSS} to match exactly the heterotic one-loop
couplings (with the appropriate Wilson lines switched on)~! Similarly,
by summing up $\Cp4$ exchange between any given brane and all the other
ones, the contribution to $\tr F_{SO(8)_1}^2\tr R^2$ is given by
$\sum_{i=2}^4\Delta_{1i}(T)\sim \ln[\eta(2T)]$ -- this again matches
the heterotic result.

\section{Global Interactions from cyclic $D$-brane charges}

There are moduli-dependent corrections also to other eight-form terms
in the effective action, eg., to $(\tr F^2)^2$ and $\tr R^4$
(again, we will focus here only on the perturbative, potentially
singular contributions). In the usually considered situation, where
one focuses on pairs of $D$- or orientifold-branes
\cite{DJM,MSM,OHirz}, such terms arise from integrating out
$\Cp0\!-\!\Cp8$ exchange between the two branes, each equipped with
couplings like $\LCS=Q_7\cdot \Cp8+\dots+\Cp0\wedge Y_8(F,R)$, where
$Y_8$ is some 8-form. However, in the present context, the 7-brane
charge is cancelled locally on every plane so that $Q_7\equiv0$. This
means that naive $\Cp0\!-\!\Cp8$ exchange cannot contribute to these
couplings, a fact that has been slightly mis-stated in
ref.~\cite{WLSS}.  But how do these moduli-dependent corrections then
arise~?

The point is that despite our 7-planes do not have net $\ZZ$-valued
$D$-brane charge (no logarithmic monodromy), there is still a remnant
left, which is reflected by the finite order $\ZZ_N$ monodromies. It is
this ``torsion'' piece of the $D$-brane charge that is responsible for
the requisite long-range interactions. This can be seen by analyzing
the interactions in terms of string junctions \cite{junct} (with the
appropriate quantum numbers so as to contribute to the terms in
question). Similar to what is familiar from orientifold planes,  what
one finds are string trajectories that loop around other planes and do
not couple to them via the local Chern-Simons terms in \eq{D4CS}; see
Fig.2.

\goodbreak
\figinsert{global}{
Interactions probing a $\ZZ_3$ torsion piece of $D$-brane charge. On the
left, a string junction is shown that contributes to $({\tr
F_{SU(3)}}^2)^2$, while the one on the right contributes to $\tr R^4$.
We also show how they lift to periods on the covering curve. The
junctions give rise to logarithmic singularities when the planes
collide (the actual BPS geodesics may look quite different, though, and
might involve prongs.)
}{2.0in}{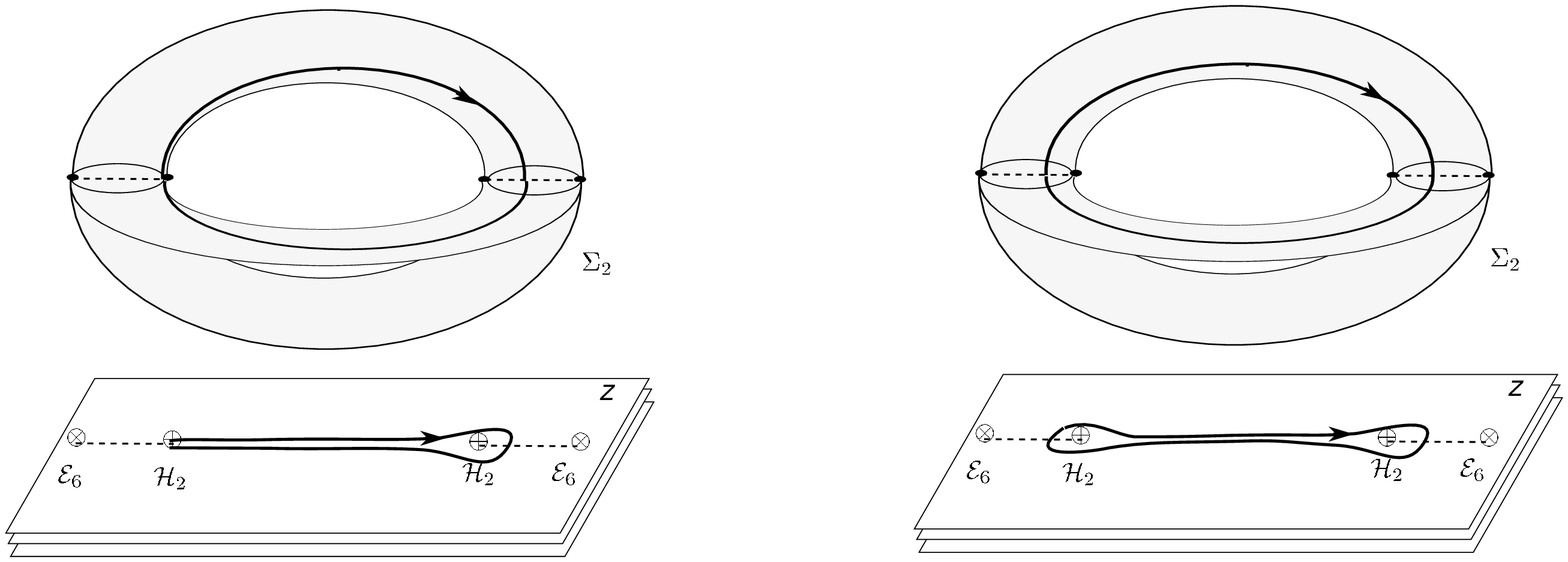}

The question arises to what extent it is possible to represent such
interactions in terms of local couplings on the plane world-volumina.
One might imagine pulling the string loop through the plane location,
thereby creating a prong that ends on the plane and thus might have a
local coupling to it. Indeed for $\cD_4$ 7-planes, which are related to
$\ZZ_2$ orientifold planes, the interaction can be treated in conformal
field theory, where it can be represented by a cross cap vertex
localized on the plane. On the other hand, for the $\ZZ_N$
generalizations listed in Table 1, there is no conformal field theory
description because these 7-planes are associated with strong (and
frozen) type IIB coupling, $\taus$. It thus seems more natural to view
these interactions not in terms of localized couplings, but rather as
analogous to those of ``Alice strings'' \cite{alice}, which do have
long-range interactions but no locally defined charge density.

At any rate, the heterotic string computation tells exactly what the
answer must be, and it turns out that we obtain the correct result if
we simply take the same Green's functions as before. That is, adding up
all the relevant exchanges, we obtain $\sum_{i=2}^4\Delta_{1i}(T)\sim
\ln[\eta(2T)](\tr F_{SO(8)_1}^2)^2$, and the same functional form as
well for $(\tr R^2)^2$ and $\tr R^4$; up to normalization, this indeed
reproduces the heterotic one-loop couplings. Note that the closed loops
in the $z$-plane that contribute to the gravitational couplings lift to
elliptic curves in the $K3$, which is consistent with the expectation
that gravitational couplings are not corrected at type IIB closed
string tree level.

\goodbreak \figinsert{Dinter}{Comparison of some of the $T$-dependent
interactions between $\cD_4$ planes in the $SO(8)^4$ model. Shown on
the left are the $\Cp4$ exchanges that couple locally and lead to the
couplings $\tr F_{SO(8)_1}^2\tr F_{SO(8)_j}^2$. In the middle we see
contributions to $(\tr F_{SO(8)_1}^2)^2$, and on the right some of the
contributions to $\tr R^4$, the dashed lines denoting $\ZZ_2$ branch
cuts.}{1.4in}{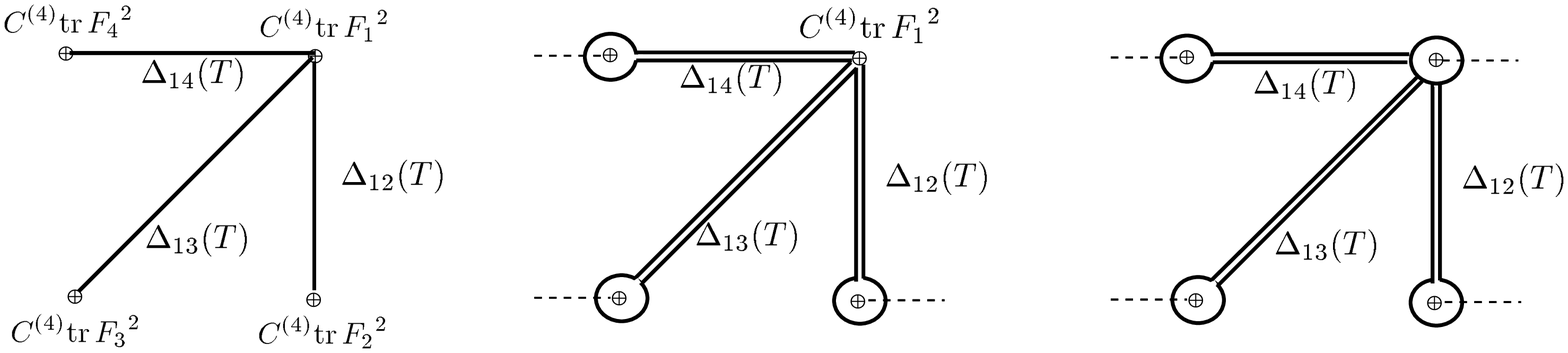}

We hope that the above considerations will be relevant in a
broader context, perhaps in the $K$-theoretic approach \cite{Ktheory}
to generalized $D$-brane charges, where indeed $\ZZ_N$ charges play an
important role. We may also present more detailed results elsewhere.


\vskip0.5cm
\goodbreak
\noindent
{\large \bf Acknowledgements}
\smallskip
\noindent
We thank Nick Warner for collaboration on this subject and Peter Mayr
for discussions, and moreover W.L.\ thanks the organizers of the
Symposium for a pleasant workshop.

\newcommand\nil[1]{{}}
\newcommand\nihil[1]{{\sl #1}}
\newcommand\ex[1]{}
\def\eprt#1{{\tt #1}}
\newcommand\br{{\hfill\break}}

\newcommand{\np}[3]{{\em Nucl.\ Phys.}\ {B#1#2#3}\ }

\hfuzz=20pt



\lref\Fth{
{C.\ Vafa,
 \nihil{Evidence for F-theory,}
 Nucl.\ Phys.\ {\bf B469} (1996) 403-418,
 \eprt{hep-th/9602022}.}
}

\lref\HM{
{J.\ Harvey and G.\ Moore,
 \nihil{Algebras, BPS States, and Strings,}
 Nucl.\  Phys.\ {\bf B463} (1996) 315-368,
 \eprt{hep-th/9510182}.}}

\lref \DKLII{L. Dixon, V. Kaplunovsky and J. Louis,
 \nihil{Moduli dependence of string loop corrections to gauge coupling
constants,} Nucl.\ Phys.\ {\bf B355} (1991) 649-688.}

\lref\Sen{
{A.\ Sen, \nihil{F-theory and Orientifolds,}\np {\bf 475} (1996) 562
 \eprt{hep-th/9605150}.}}

\lref\BK{C.\ Bachas and E.\ Kiritsis,
 \nihil{$F^4$ terms in N=4 string vacua,}
 Nucl.\  Phys.\  Proc.\  Suppl.\ {\bf 55B} (1997) 194,
 \eprt{hep-th/9611205}.}

\lref\BFKOV{
{C.\ Bachas, C.\ Fabre, E.\ Kiritsis, N.\ Obers and P.\ Vanhove,
 \nihil{Heterotic/type I duality and D-brane instantons,}
 Nucl.\  Phys.\ {\bf B509} (1998) 33,
 \eprt{hep-th/9707126}.}
}

\lref\typeI{
{C.\ Bachas,
 \nihil{Heterotic versus type-I,}
 Nucl.~ Phys.~ Proc.~ Suppl.~{\bf 68} (1998) 348,
 \eprt{hep-th/9710102};}
\br
{M.\ Bianchi, E.\ Gava, F.\ Morales and K.\ S.\ Narain,
 \nihil{D strings in unconventional type I vacuum configurations,}
 \eprt{hep-th/9811013}.}
\br
{E.\ Gava, A.\ Hammou, J.\ F.\ Morales and K.\ S.\ Narain,
 \nihil{On the perturbative corrections around D string instantons,}
 \eprt{hep-th/9902202.}
}
\br
{K.\ Foerger and S.\ Stieberger,
 \nihil{Higher derivative couplings and heterotic
   type I duality in eight-dimensions,}
 \eprt{hep-th/9901020.}
}
\br
{M.\ Gutperle,
 \nihil{A Note on heterotic/type I' duality and D0 brane quantum
mechanics,}
 \eprt{hep-th/9903010}.}
}

\lref\OHirz{
{B.\ Craps and F.\ Roose,
 \nihil{Anomalous D-brane and orientifold couplings from the boundary
state,}
 Phys.~ Lett.~{\bf B445} (1998) 150,
 \eprt{hep-th/9808074};}
\br
{B.\ Stefanski,
 \nihil{Gravitational couplings of D-branes and O-planes,}
 \eprt{hep-th/9812088}.}
}

\lref\MSM{J.\ Morales, C.\ Scrucca and M.\ Serone,
 \nihil{Anomalous couplings for D-branes and O-planes,}
 \eprt{hep-th/9812071}.}

\lref\GHM{M.\ Green, J.\ Harvey and G.\ Moore,
 \nihil{I-brane inflow and anomalous couplings on d-branes,}
 Class.~ Quant.~ Grav.~{\bf 14} (1997) 47-52,
 \eprt{hep-th/9605033}.}

\lref\KDSM{K.\ Dasgupta and S.\ Mukhi,
 \nihil{F-theory at constant coupling,}
 Phys.\  Lett.\ {\bf B385} (1996) 125-131,
 \eprt{hep-th/9606044}.}

\lref\DJM{K.\ Dasgupta, D.\ Jatkar and S.\ Mukhi,
 \nihil{Gravitational couplings and $\ZZ_2$ orientifolds,}
 Nucl.~ Phys.~{\bf B523} (1998) 465,
 \eprt{hep-th/9707224}.}

\lref\ko{E.\ Kiritsis and N.\ Obers,  \nihil{Heterotic type I duality
in $d <10$-dimensions, threshold corrections and D-instantons,} {\it
JHEP} {\bf 10} (1997) 004, \eprt {hep-th/9709058}.
}

\lref\WL{W. Lerche, {
 \nihil{Elliptic index and superstring effective actions,}
 Nucl.\  Phys.\ {\bf B308} (1988) 102.}}

\lref\ellg {A.\ Schellekens and N.\ Warner,
{\nihil{Anomalies, characters and strings,}
 Nucl.\  Phys.\ {\bf B287} (1987) 317;}\br
{E.\ Witten,
 \nihil{Elliptic genera and quantum field theory,}
 Comm.\  Math.\  Phys.\ {\bf 109} (1987) 525;}\br
W. Lerche, B.E.W. Nilsson, A.N. Schellekens and N.P. Warner,
 Nucl.\  Phys.\  {\bf 299} (1988) 91.}

\lref\junct{ see eg.:\br
{M.\ Gaberdiel, T.\ Hauer and B.\ Zwiebach,
 \nihil{Open string-string junction transitions,}
 Nucl.~ Phys.~{\bf B525} (1998) 117,
 \eprt{hep-th/9801205};}
\br
{O.\ DeWolfe and B.\ Zwiebach,
 \nihil{String junctions for arbitrary Lie algebra representations,}
 Nucl.~ Phys.~{\bf B541} (1999) 509,
 \eprt{hep-th/9804210}.}
}

\lref\alice{A.\ S.\ Schwarz,
 \nihil{Field theories with no local conservation of the electric
charge,}
 Nucl.~ Phys.~{\bf B208} (1982) 141.}

\lref\WLSS{W.\ Lerche and S.\ Stieberger,
 \nihil{Prepotential, mirror map and F-theory on K3,}
 \eprt{hep-th/9804176}, to appear in ATMP 2.5.}

\lref\LSWA{W.\ Lerche, S.\ Stieberger and N.\ Warner,
 \nihil{Quartic gauge couplings from K3 geometry,}
 \eprt{hep-th/9811228}.}

\lref\LSWB{W.\ Lerche, S.\ Stieberger and N.\ Warner,
 \nihil{Prepotentials from symmetric products,}
 \eprt{hep-th/9901162}.}

\lref\Ktheory{E.\ Witten,
 \nihil{D-branes and K theory,}
 \eprt{hep-th/9810188}.}

\lref\CB{C.\ Bachas,
 \nihil{(Half) a lecture on D-branes,}
 \eprt{hep-th/9701019}.}

\lref\Kod{K.\ Kodaira, Ann.\ Math.\ {\bf 77} (1963) 563;
Ann.\ Math.\ {\bf 78} (1963) 1.}


\end{document}